\PassOptionsToPackage{unicode}{hyperref}
\PassOptionsToPackage{hyphens}{url}
\PassOptionsToPackage{dvipsnames,svgnames,x11names}{xcolor}
\documentclass[
]{article}
\usepackage{amsmath,amssymb}
\usepackage{lmodern}
\usepackage{iftex}

\usepackage[T1]{fontenc}
\usepackage[utf8]{inputenc}
\usepackage{textcomp} 

\IfFileExists{upquote.sty}{\usepackage{upquote}}{}
\IfFileExists{microtype.sty}{
  \usepackage[]{microtype}
  \UseMicrotypeSet[protrusion]{basicmath} 
}{}
\makeatletter
\@ifundefined{KOMAClassName}{
  \IfFileExists{parskip.sty}{%
    \usepackage{parskip}
  }{
    \setlength{\parindent}{0pt}
    \setlength{\parskip}{6pt plus 2pt minus 1pt}}
}{
  \KOMAoptions{parskip=half}}
\makeatother
\usepackage{xcolor}
\NewDocumentCommand\citeproctext{}{}
\NewDocumentCommand\citeproc{mm}{%
  \begingroup\def\citeproctext{#2}\cite{#1}\endgroup}
\makeatletter
 \let\@cite@ofmt\@firstofone
 \def\@biblabel#1{}
 \def\@cite#1#2{{#1\if@tempswa , #2\fi}}
\makeatother
\newlength{\cslhangindent}
\newlength{\csllabelwidth}
\newenvironment{CSLReferences}[2] 
 {\begin{list}{}{%
  \setlength{\itemindent}{0pt}
  \setlength{\leftmargin}{0pt}
  \setlength{\parsep}{0pt}
  \ifodd #1
   \setlength{\leftmargin}{\cslhangindent}
   \setlength{\itemindent}{-1\cslhangindent}
  \fi
  \setlength{\itemsep}{#2\baselineskip}}}
 {\end{list}}
\usepackage{calc}

\ifLuaTeX
\usepackage[bidi=basic]{babel}
\else
\usepackage[bidi=default]{babel}
\fi
\babelprovide[main,import]{american}

\def\languageshorthands#1{}
\ifLuaTeX
  \usepackage{selnolig}  
\fi
\IfFileExists{bookmark.sty}{\usepackage{bookmark}}{\usepackage{hyperref}}
\IfFileExists{xurl.sty}{\usepackage{xurl}}{} 
\hypersetup{
  pdftitle={OrbDot: A Python package for studying the secular evolution
of exoplanet orbits},
  pdfauthor={Simone R. Hagey, Aaron Boley},
  pdflang={en-US},
  colorlinks=true,
  linkcolor={Maroon},
  filecolor={Maroon},
  citecolor={Blue},
  urlcolor={Blue},
  pdfcreator={LaTeX via pandoc}}

\title{OrbDot: A Python package for studying the secular evolution of
exoplanet orbits}

\definecolor{c53baa1}{RGB}{83,186,161}
\definecolor{c202826}{RGB}{32,40,38}

\usepackage[affil-it]{authblk}
\usepackage{orcidlink}
\author[1%
  ]{Simone R. Hagey%
    \,\orcidlink{0000-0001-8072-0590}\,%
    }
\author[1%
  ]{Aaron Boley%
    \,\orcidlink{0000-0002-0574-4418}\,%
    }

\affil[1]{Department of Physics and Astronomy, The University of British
Columbia, 6224 Agricultural Road Vancouver, BC, V6T 1Z1, Canada%
  }
\date{1 September 2025}

\begin{document}
\maketitle

\section{Summary}\label{summary}

Gradual changes in exoplanet orbits, known as secular variations, can be
detected through observations of transits, eclipses, and radial
velocities that span multiple decades in time. Their detection and
characterization enable the study of a wide range of dynamical
phenomena, such as orbital decay and precession, which operate on
timescales of millions of years. Under certain conditions, measurements
of secular variations can even probe the interior structure of
exoplanets, providing a unique tool for understanding exoplanet
formation and evolution.

The necessity to search over many orbital epochs coupled with an
ever-growing archive of exoplanet observations creates a need for fast
and flexible open-source software that can reliably detect gradual
changes in exoplanet orbits. \texttt{OrbDot} addresses this need by
offering robust tools for fitting secular evolution models to exoplanet
transit and eclipse mid-times, transit durations, and radial velocity
data.

A key advantage of \texttt{OrbDot} is its ability to fit multiple types
of data simultaneously, which can help to break parameter degeneracies.
It also excels at assisting in result interpretation by generating
reports on model comparisons and assessments of various physical effects
in the context of the models and their corresponding theory. For
example, analysis reports could determine key parameters for assessing
tidal energy dissipation, apsidal precession mechanisms, variations due
to systemic proper motion, and the dynamical effects of non-resonant
companion objects, depending on the applied models.

\texttt{OrbDot} remains highly efficient with multiple data types and a
high number of free parameters, as it utilizes powerful nested sampling
algorithms of the \texttt{nestle} (\citeproc{ref-nestle}{Barbary, 2021};
\citeproc{ref-Skilling:2006}{Skilling, 2006}) and PyMultiNest
(\citeproc{ref-Buchner:2014}{Buchner et al., 2014};
\citeproc{ref-Feroz:2009}{Feroz et al., 2009}) packages. The intricacies
of the implementation are abstracted such that the \texttt{OrbDot} input
files are simple and the method calls require only a list of free
parameters, along with the desired model for fitting.

Extensive documentation, including examples, is hosted on
\href{https://orbdot.readthedocs.io}{ReadTheDocs}.

The examples demonstrate that \texttt{OrbDot} can quickly reproduce
literature results using only a few lines of code. Readers may be
especially interested in the \texttt{OrbDot} example analysis of the
transit and eclipse mid-times of Hot Jupiter WASP-12 b, which is
well-known for showing strong evidence for orbital decay.

A complementary case study of TrES-1 b (\citeproc{ref-Hagey:2025}{Hagey
et al., 2025}) illustrates the full capabilities of \texttt{OrbDot},
placing it in a broader scientific context. Moreover, an early version
of this code was used for the orbital analysis of the Hot Neptune
LTT-9779 b, published in Edwards et al.
(\citeproc{ref-Edwards:2023}{2023}).

\section{Statement of need}\label{statement-of-need}

Many exoplanet systems now have transit and radial velocity data
spanning over a decade, enabling studies of secular variations. While
tools for analyzing short-term transit variations exist, there is a lack
of open-source software dedicated to long-term orbital evolution. This
does not, however, reflect a lack of interest, as the number of such
studies is growing rapidly.

\texttt{OrbDot} lowers the barrier to entry for researchers at all
levels, including undergraduates, by making advanced statistical methods
accessible without requiring extensive computational experience. Despite
its ease of use, \texttt{OrbDot} is not intended to be a black box.
Rather, with extensive documentation, examples, and accessible source
code, it is presented to the community with transparency that lends
itself to community contributions and independent verification of
results. It is designed to be easily extended, as the nested sampling
framework supports custom log-likelihood models with free parameters
that are part of the \texttt{OrbDot} ecosystem. This ensures that the
software may evolve to meet the needs of the research community.

\section{Similar software}\label{similar-software}

Some existing tools have features that overlap with \texttt{OrbDot}'s
functionality, but none provides its full suite of capabilities. The
most similar codes, \texttt{Susie} (\citeproc{ref-Barker:2024}{Barker et
al., 2024}; \citeproc{ref-susie}{Barker \& Kirk, 2025}) and
\texttt{PdotQuest} (\citeproc{ref-Wang:2024}{Wang et al., 2024}), which
are not fully packaged software, are designed to fit secular evolution
models to transit and eclipse timing data, but not to radial velocities.
\texttt{OrbDot} also has greater flexibility in model fitting than
\texttt{Susie} and \texttt{PdotQuest}. For example, \texttt{Susie}
employs simple least-squares fitting, and while \texttt{PdotQuest}
(\citeproc{ref-Wang:2024}{Wang et al., 2024}) uses MCMC, it currently
supports only the orbital decay model, making both codes narrower in
scope than \texttt{OrbDot}. Moreover, \texttt{OrbDot} includes tools for
theoretical interpretation.

The well-known package \texttt{TTVFast} (\citeproc{ref-Deck:2014}{Deck
et al., 2014}) is highly robust and capable of modeling both transit and
radial velocity data, but it is focused on short-term timing variations
driven by multi-planet dynamics near mean-motion resonances.
\texttt{OrbDot}, in contrast, explores long-term secular models.
Similarly, general-purpose frameworks such as \texttt{juliet}
(\citeproc{ref-Espinoza:2019}{Espinoza et al., 2019}),
\texttt{exoplanet} (\citeproc{ref-ForemanMackey:2021}{Foreman-Mackey et
al., 2021}), \texttt{EXOFAST} (\citeproc{ref-Eastman:2019}{Eastman et
al., 2019}), \texttt{ExoStriker} (\citeproc{ref-Trifonov:2019}{Trifonov,
2019}), and \texttt{allesfitter} (\citeproc{ref-Gunther:2021}{Günther \&
Daylan, 2021}) can jointly model transit and RV data, but their TTV
models are also restricted to short-term variations, with an emphasis on
transit light curve fitting rather than directly using transit and
eclipse mid-times to constrain models.

The RV-focused packages \texttt{RadVel}
(\citeproc{ref-Fulton:2018}{Fulton et al., 2018}) and \texttt{Kima}
(\citeproc{ref-Faria:2018}{Faria et al., 2018}) provide flexible
modeling of radial velocity datasets, but do not have a framework for
incorporating transit and eclipse data into the modeling. The codes also
have limited options for studying long-term trends. For example,
\texttt{RadVel} (\citeproc{ref-Fulton:2018}{Fulton et al., 2018})
includes linear and quadratic RV terms, which \texttt{OrbDot} also
supports, but it does not model evolving orbital elements. \texttt{Kima}
(\citeproc{ref-Faria:2018}{Faria et al., 2018}) incorporates an apsidal
precession model, but only for circumbinary planets -- a niche
application.

In summary, \texttt{OrbDot} is a fully packaged, documented, and
maintained software suite designed specifically for studies of secular
orbital evolution. It unifies transit, eclipse, and RV data, bypasses
light-curve fitting to focus directly on timing measurements, implements
robust Bayesian inference with nested sampling, and provides a flexible
framework for selecting models, priors, and parameterizations. In
addition, it integrates tools for theoretical analysis, making
\texttt{OrbDot} the first software to systematically support
data-driven, comprehensive studies of long-term exoplanet orbital
dynamics.

\section{Acknowledgements}\label{acknowledgements}

This work was supported, in part, by an NSERC Discovery Grant
(DG-2020-04635) and the University of British Columbia. SH's
contribution was further supported, in part, by an NSERC PGS-D and a Li
Tze Fong Fellowship.

\section*{References}\label{references}
\addcontentsline{toc}{section}{References}

\phantomsection\label{refs}

\end{document}